\documentclass[showpacs,aps,graphicx,twocolumn]{revtex4}%and
\usepackage{graphicx}

\begin{document}

\title{High-efficient two-step entanglement purification using hyperentanglement}

\author{Lan Zhou,$^{1}$ Yu-Bo Sheng$^{2}$\footnote{shengyb@njupt.edu.cn} }
\address{$^1$ School of Science, Nanjing University of Posts and Telecommunications, Nanjing,
210003, China\\
$^2$Institute of Quantum Information and Technology, Nanjing University of Posts and Telecommunications, Nanjing, 210003,  China\\}

\date{\today }
\begin{abstract}
Entanglement purification is a powerful method to distill the high-quality entanglement from low-quality entanglement. In the paper, we propose an efficient two-step entanglement purification protocol (EPP) for the polarization entanglement by using only one copy of two-photon hyperentangled state in polarization, spatial-mode, and time-bin DOFs. We suppose that the entanglement in all DOFs suffer from channel noise. In two purification steps, the parties can reduce the bit-flip error and phase-flip error in polarization DOF by consuming the imperfect entanglement in the spatial-mode and time-bin DOFs, respectively. This EPP effectively reduces the consumption
of entanglement pairs and the experimental difficulty. Moreover, if consider the practical photon transmission and detector efficiencies, our EPP has much higher purification efficiency than previous recurrence EPPs. Meanwhile, when one or two purification steps fail, the distilled mixed state may have residual entanglement. Taking use of the residual entanglement, the parties may still distill higher-quality polarization entanglement. Even if not, they can still reuse the residual entanglement in the next purification round. The existence of residual entanglement benefits for increasing the yield of the EPP. All the above advantages
make our EPP have potential application in future quantum information processing.
\end{abstract}
 \maketitle
\section{Introduction}
Entanglement is an indispensable resource which is widely applied in quantum communication field, such as
quantum teleportation\cite{teleportation1,teleportation2,teleportation3}, quantum repeater \cite{repeater1,repeater2,repeater3},  quantum key distribution (QKD) \cite{qkd}, quantum secret sharing (QSS)
\cite{qss}, and quantum secure direct communication (QSDC) \cite{qsdc1,qsdc2,qsdc3,qsdc4}. The above applications often require the maximal entanglement. However, entanglement is generally fragile due to the channel noise. During the practical applications, the degraded entanglement may decrease the quantum communication efficiency or even make quantum communication insecure.

Entanglement purification which was first proposed by Bennett \emph{et al.} in 1996 \cite{EPP0} is an efficient method to distill high quality entanglement from low quality entanglement with local operation and classical communications (LOCC). Recurrence entanglement purification is the most common entanglement purification form, which has been well developed in both theory and experiment \cite{EPP0,Deustch,Murao,Pan1,Pan2,Pan3,graph,atom,Cheong,sheng1,sheng2,Zhou1,Wang,Zhou2,Zhang,Dur,Rozpeedek,Krastanov,Wu,zhouap,zhouoe,hu,Du,nest,network,shenghyper1,shenghyper2}. The recurrence entanglement purification protocols (EPPs) require two or more copies of low-quality entangled states from the same enables. After two communication parties in distant locations performing the controlled-not (CNOT) or other similar operations, one pair of low-quality entangled state is measured. If the purification is successful, the fidelity of left photon pair can be increased. For example, in 2001, Pan \emph{et al.} presented an EPP of general mixed entangled states with linear optical elements \cite{Pan1}, and later they improved their EPP by adopting available parametric down conversion sources \cite{Pan2}. In 2003, Pan \emph{et al.} demonstrated the experiment of the entanglement purification for general mixed states of polarization-entangled photons \cite{Pan3}. In 2008, Sheng \emph{et al.} proposed an EPP based on nondestructive quantum nondemolition
detectors \cite{sheng1}. In 2017, Pan \emph{et al.} experimentally realized the nested purification for a linear optical quantum repeater \cite{nest}. In addition, the experimental purification between two-atom entanglement and solid state quantum network nodes were also demonstrated \cite{atom,network}. In 2010, Sheng \emph{et al.} proposed the deterministic EPPs (DEPPs) by adopting the hyperentanglement \cite{shenghyper1,shenghyper2}.

Although the recurrence entanglement purification has been well studied, existing recurrence EPPs often have relatively low yield. The reason is that in each purification round, at least one pair of low-quality entangled states should be consumed. In practical applications, entanglement
purification process often has to be iterated for many rounds to obtain high-fidelity entangled pairs, so that a large amount of low-quality entangled
pairs have to be consumed. It is a big waste of the precious entanglement resources. In 2021, Hu \emph{et al.} proposed and experimentally demonstrated the
first long-distance polarization entanglement recurrence purification using only one polarization-spatial-mode hyperentangled photon pair \cite{hu}. They supposed that the entanglement in both polarization and spatial-mode DOFs suffer from one kind of error, say, the bit-flip error or phase-flip error. After performing the EPP, they obtained a significant improvement in the fidelity of polarization entanglement. This protocol effectively reduces consumption of copies of entanglement pairs, especially in purification consisting of many rounds.

Actually, in practical entanglement distribution process, the bit-flip error and phase-flip error may be occurred simultaneously. In the paper, we consider a more general recurrence EPP which can simultaneously reduce the bit-flip error and phase-flip error of the polarization entanglement by using only one pair of polarization-spatial-time-bin hyperentangled photon pair. We choose the spatial mode and time-bin entanglement for the entanglement in both two DOFs, especially in time-bin DOF being highly robust to the channel noise. The entanglement in time-bin DOF has been successfully used in
the transmission of qubits over hundreds of kilometers \cite{Pan2,time1,time2,time3,time4} and in teleportation using real-world fiber networks \cite{time5,time6}. In 2005, Barreiro \emph{et al.} experimentally demonstrated
the generation of hyperentanglement in polarization, spatial-mode and time-energy
DOFs of photon systems using pairs of photons produced in spontaneous parametric down-conversion \cite{generation5}. In our protocol, we suppose that the entanglement in all DOFs suffer from channel noise and degrade to mixed states. As the entanglement in above three DOFs have different noise robustness, after the photon transmission, the fidelities in three DOFs are naturally different. After performing our EPP, we can efficiently reduce both the bit-flip error and phase-flip error rate in polarization DOF by consuming the imperfect entanglement in the spatial-mode and time-bin DOFs. Moreover, we will prove that if a purification step fails, there may exist residual entanglement in the corresponding distilled mixed state. By using the residual entanglement, we may also increase the fidelity of the polarization entanglement after the whole purification process.

The paper is organized as follows. In Sec. II, we describe our EPP in a simple case where the entanglement in spatial-mode and time-bin DOFs only suffer from a bit-flip error. In Sec. III, we extend our EPP to a general case where both the entanglement in spatial-mode and time-bin DOFs suffer from both bit-flip error and phase-flip error. In Sec. IV, we make a discussion. In Sec. V, we make a conclusion.

\section{Entanglement purification principle}
In this section, we propose  our long-distance EPP using a hyperentangled photon pair. Suppose the photon hyperentanglement source S generates a two-photon hyperentanglement in polarization, spatial-mode, and time-bin DOFs, which can be described as
\begin{eqnarray}
&&|\Phi^{+}_{p}\rangle|\Phi^{+}_{s}\rangle|\Phi^{+}_{t}\rangle=\frac{1}{\sqrt{2}}(|HH\rangle+|VV\rangle)\nonumber\\
&\otimes&\frac{1}{\sqrt{2}}(|a_{1}'b_{1}'\rangle+|a_{2}'b_{2}'\rangle)
\otimes\frac{1}{\sqrt{2}}(|LL\rangle+|SS\rangle).\label{initial}
\end{eqnarray}
Here, $H$ ($V$) represents the horizontal (vertical) polarization, $a_{1}'$, $a_{2}'$, $b_{1}'$ and $b_{2}'$ are four different spatial modes, and $L$ ($S$) represents the long (short) time-bin.

 The photon in $a_{1}'$ and $a_{2}'$ are sent to Alice, while the photon in $b_{1}'$ and $b_{2}'$ are sent to Bob. After long distance transmission, the channel noise may degrade the entanglement in all DOFs. Here, we suppose that the entanglement in polarization DOF degrade to a Werner state with the form of
 \begin{eqnarray}
\rho_{p}&=&p_{p}|\Phi^{+}_{p}\rangle\langle\Phi^{+}_{p}|+\frac{1-p_{p}}{3}(|\Psi^{+}_{p}\rangle\langle \Psi^{+}_{p}|\nonumber\\
&+&|\Phi^{-}_{p}\rangle\langle \Phi^{-}_{p}|
+|\Psi^{-}_{p}\rangle\langle \Psi^{-}_{p}|),\label{initialp}
\end{eqnarray}
where
 \begin{eqnarray}
|\Phi^{\pm}_{p}\rangle&=&\frac{1}{\sqrt{2}}(|HH\rangle\pm|VV\rangle), \nonumber\\
 |\Psi^{\pm}_{p}\rangle&=&\frac{1}{\sqrt{2}}(|HV\rangle\pm|VH\rangle).
\end{eqnarray}
We notice that the state $|\Phi^{+}_{p}\rangle$ becomes $|\Psi^{+}_{p}\rangle$ when a bit-flip error occurs, $|\Phi^{+}_{p}\rangle$ becomes $|\Phi^{-}_{p}\rangle$ when a phase-flip error occurs, and $|\Phi^{+}_{p}\rangle$ becomes $|\Psi^{-}_{p}\rangle$ when bit-flip error and phase-flip error both occur.

 Considering the noise robustness of the entanglement in spatial-mode and time-bin DOFs are higher than that in the polarization DOF, we first focus on a simple case where the entanglement in spatial-mode and time-bin DOFs only suffer from bit-flip error. In this case, the entanglement in both DOFs degrade to
 \begin{eqnarray}
 \rho_{s}&=&p_{s}|\Phi^{+}_{s}\rangle\langle \Phi^{+}_{s}|+(1-p_{s})|\Psi^{+}_{s}\rangle\langle \Psi^{+}_{s}|,\\
 \rho_{t}&=&p_{t}|\Phi^{+}_{t}\rangle\langle \Phi^{+}_{t}|+(1-p_{t})|\Psi^{+}_{t}\rangle\langle \Psi^{+}_{t}|,
\end{eqnarray}
 where
 \begin{eqnarray}
 |\Phi^{\pm}_{s}\rangle&=&\frac{1}{\sqrt{2}}(|a_{1}b_{1}\rangle\pm|a_{2}b_{2}\rangle), \nonumber\\
 |\Psi^{\pm}_{s}\rangle&=&\frac{1}{\sqrt{2}}(|a_{1}b_{2}\rangle\pm|a_{2}b_{1}\rangle),\nonumber\\
 |\Phi^{\pm}_{t}\rangle&=&\frac{1}{\sqrt{2}}(|LL\rangle\pm|SS\rangle), \nonumber\\
 |\Psi^{\pm}_{t}\rangle&=&\frac{1}{\sqrt{2}}(|LS\rangle\pm|SL\rangle).
 \end{eqnarray}

 Therefore, the initial state in Eq. (\ref{initial}) degrades to $\rho_{p}\otimes\rho_{s}\otimes\rho_{t}$. Here, we suppose that all the fidelity $p_{p}$, $p_{s}$, and $p_{t}$ are higher than $\frac{1}{2}$.  The schematic principle of our EPP is shown in Fig. 1.  The protocol includes two steps. In the first step, Alice and Bob correct the bit-flip error in polarization DOF by consuming the entanglement in spatial-mode DOF. In the second step, they correct the phase-flip error in  polarization DOF with the help of the entanglement in time-bin DOF.

\begin{figure*}
\centering
\includegraphics[width=13cm]{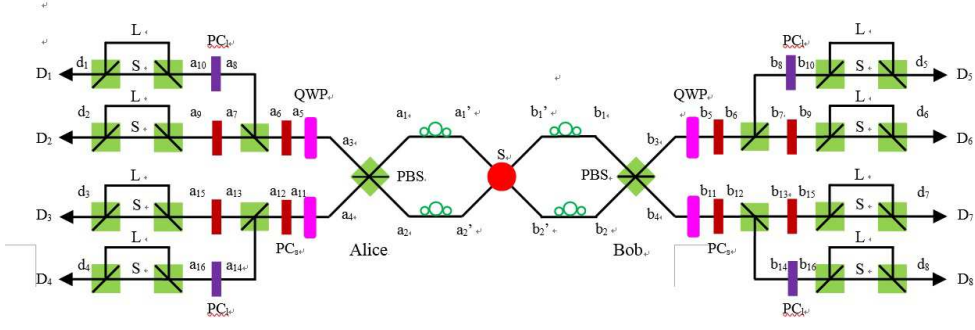}
\caption{The basic principle of our two-step EPP. Suppose a photon pair hyperentangled in polarization, spatial-mode, and time-bin DOFs suffer from channel noise, and the entanglement in all DOFs degrade to mixed states. Here, PBS means polarization beam splitter, which can totally transmit the photon in $|H\rangle$ and reflect the photon in $|V\rangle$. QWP means the $\lambda/4$-wave plate, which can make the Hadamard (H) operation in the polarization DOF. $PC_{l(s)}$ presents the Pockels cell, which can revers the polarization of a photon with the time-bin L (S). $D_{1}-D_{8}$ are the single photon detectors. }
\end{figure*}

 In the first step, we only use the states in polarization and spatial-mode DOF and leave the state in time-bin DOF unchanged, so that we first neglect the state in time-bin DOF and  only consider the mixed state as $\rho_{1}=\rho_{p}\otimes\rho_{s}$ for simplicity. There are totally eight possible cases. The photon system may be in $|\Phi^{+}_{p}\rangle|\Phi^{+}_{s}\rangle$ with the probability of $p_{p}p_{s}$, while it may be in $|\Phi^{+}_{p}\rangle|\Psi^{+}_{s}\rangle$ with the probability of $p_{p}(1-p_{s})$. Meanwhile, the photon system may be in $|\Psi^{+}_{p}\rangle|\Phi^{+}_{s}\rangle$, $|\Phi^{-}_{p}\rangle|\Phi^{+}_{s}\rangle$, or $|\Psi^{-}_{p}\rangle|\Phi^{+}_{s}\rangle$ with the equal probability of $\frac{(1-p_{p})p_{s}}{3}$, and it may be in the state $|\Psi^{+}_{p}\rangle|\Psi^{+}_{s}\rangle$, $|\Phi^{-}_{p}\rangle|\Psi^{+}_{s}\rangle$, or $|\Psi^{-}_{p}\rangle|\Psi^{+}_{s}\rangle$ with the probability of $\frac{(1-p_{p})(1-p_{s})}{3}$.

 Alice and Bob pass the photons in $a_{1}$ and $a_{2}$, $b_{1}$ and $b_{2}$ spatial modes through two polarization beam splitters (PBSs), which can totally transmit the photon in $|H\rangle$ and reflect the photon in $|V\rangle$. If the initial photon state is $|\Phi^{\pm}_{p}\rangle|\Phi^{+}_{s}\rangle$, after the PBS, the state will evolve to
 \begin{eqnarray}
 &&|\Phi^{\pm}_{p}\rangle|\Phi^{+}_{s}\rangle=\frac{1}{\sqrt{2}}(|HH\rangle\pm|VV\rangle)\otimes\frac{1}{\sqrt{2}}(|a_{1}b_{1}\rangle+|a_{2}b_{2}\rangle)\nonumber\\
 &\rightarrow&\frac{1}{\sqrt{2}}(|HH\rangle\pm|VV\rangle)\otimes\frac{1}{\sqrt{2}}(|a_{3}b_{3}\rangle+|a_{4}b_{4}\rangle).\label{p1}
\end{eqnarray}
When the initial state is $|\Psi^{\pm}_{p}\rangle|\Psi^{+}_{s}\rangle$, it will evolve to
\begin{eqnarray}
 &&|\Psi^{\pm}_{p}\rangle|\Psi^{+}_{s}\rangle=\frac{1}{\sqrt{2}}(|HV\rangle\pm|VH\rangle)\otimes\frac{1}{\sqrt{2}}(|a_{1}b_{2}\rangle+|a_{2}b_{1}\rangle)\nonumber\\
 &\rightarrow&\frac{1}{\sqrt{2}}(|HV\rangle\pm|VH\rangle)\otimes\frac{1}{\sqrt{2}}(|a_{3}b_{3}\rangle+|a_{4}b_{4}\rangle).\label{p2}
\end{eqnarray}
 All the four above states make the spatial modes $a_{3}b_{3}$ or $a_{4}b_{4}$ each have one photon. In these cases, the first step is successful.  On the other hand, if the initial state is one of the other four cases, say, $|\Phi^{\pm}_{p}\rangle|\Psi^{+}_{s}\rangle$ and $|\Psi^{\pm}_{p}\rangle|\Phi^{+}_{s}\rangle$, after the PBSs, we can obtain the spatial-modes $a_{3}b_{4}$ or $a_{4}b_{3}$ each has one photon and the first step fails.

 As a result, when the first step is successful, we can distill a new mixed state as
 \begin{eqnarray}
 \rho_{1p}&=&F_{1}|\Phi^{+}_{p}\rangle\langle\Phi^{+}_{p}|+F_{2}|\Phi^{-}_{p}\rangle\langle\Phi^{-}_{p}|+F_{3}(|\Psi^{+}_{p}\rangle\langle\Psi^{+}_{p}|\nonumber\\
 &+&|\Psi^{-}_{p}\rangle\langle\Psi^{-}_{p}|),\label{ro1}
 \end{eqnarray}
 in the spatial modes $a_{3}b_{3}$ or $a_{4}b_{4}$ with the success probability of
 \begin{eqnarray}
 P_{1}&=&p_{p}p_{s}+\frac{1-p_{p}}{3}[p_{s}+2(1-p_{s})]\nonumber\\
 &=&\frac{1}{3}(4p_{p}p_{s}-p_{s}-2p_{p}+2).
 \end{eqnarray}
 The four coefficients in Eq. (\ref{ro1}) can be written as
\begin{eqnarray}
F_{1}&=&\frac{p_{p}p_{s}}{P_{1}}=\frac{3p_{p}p_{s}}{4p_{p}p_{s}-p_{s}-2p_{p}+2},\nonumber\\
F_{2}&=&\frac{(1-p_{p})p_{s}}{3P_{1}}=\frac{(1-p_{p})p_{s}}{4p_{p}p_{s}-p_{s}-2p_{p}+2},\nonumber\\
F_{3}&=&\frac{(1-p_{p})(1-p_{s})}{3P_{1}}=\frac{(1-p_{p})(1-p_{s})}{4p_{p}p_{s}-p_{s}-2p_{p}+2}.\label{f1}
\end{eqnarray}
It is obvious that when $p_{p}>\frac{1}{2}$ and $p_{s}>\frac{1}{2}$, the rate of $|\Psi^{\pm}_{p}\rangle$ ($F_{3}$) is smaller than their original rate $\frac{(1-p_{p})}{3}$. As a result, the first step can  reduce the rate of $|\Psi^{\pm}_{p}\rangle$.
Moreover, the reduction of bit-flip error directly increases the fidelity of $|\Phi^{+}_{p}\rangle$. We can obtain $F_{1}>p_{p}$ and $F_{1}>p_{s}$ when $p_{p}>\frac{1}{2}$ and $\frac{5p_{p}-2}{4p_{p}-1}>p_{s}>\frac{1}{2}$. However, the first step cannot deal with the phase-flip error ($|\Phi^{-}_{p}\rangle$) and the rate of $|\Phi^{-}_{p}\rangle$ is still in a relatively high level. Next, we try to correct the phase-flip error.

In the second step, we require to consume the entanglement in the time-bin DOF. After the first step, the whole photon system collapse to $\rho_{1p}\otimes\rho_{t}$ in $a_{3}b_{3}$ or $a_{4}b_{4}$ modes. We first consider that the photons are in $a_{3}b_{3}$. Alice and Bob first pass the photons in $a_{3}b_{3}$ modes through two $\lambda/4$-wave plates (QWPs), respectively. The QWP performs a Hadamard (H) operation in the polarization DOF, which makes $|H\rangle\rightarrow\frac{1}{\sqrt{2}}(|H\rangle+|V\rangle)$ and $|V\rangle\rightarrow\frac{1}{\sqrt{2}}(|H\rangle-|V\rangle)$. After the H operation,
$|\Phi^{-}_{p}\rangle\leftrightarrow|\Psi^{+}_{p}\rangle$, while $|\Phi^{+}_{p}\rangle$ and $|\Psi^{-}_{p}\rangle$ keep unchanged. In this way, they can transform  $\rho_{1p}$ to
\begin{eqnarray}
\rho_{2p}&=&F_{1}|\Phi^{+}_{p}\rangle\langle\Phi^{+}_{p}|+F_{2}|\Psi^{+}_{p}\rangle\langle\Psi^{+}_{p}|+F_{3}(|\Phi^{-}_{p}\rangle\langle\Phi^{-}_{p}|\nonumber\\
&+&|\Psi^{-}_{p}\rangle\langle \Psi^{-}_{p}|)
\end{eqnarray}
in the spatial modes $a_{5}b_{5}$. The whole photon system $\rho_{2p}\otimes\rho_{t}$ can be described as follows. With probability of $F_{1}p_{t}$ and $F_{1}(1-p_{t})$ the photon pair is in $|\Phi^{+}_{p}\rangle|\Phi^{+}_{t}\rangle$ and $|\Phi^{+}_{p}\rangle|\Psi^{+}_{t}\rangle$, respectively. With a probability of  $F_{2}p_{t}$ and $F_{2}(1-p_{t})$, it is in $|\Psi^{+}_{p}\rangle|\Phi^{+}_{t}\rangle$ and $|\Psi^{+}_{p}\rangle|\Psi^{+}_{t}\rangle$, respectively. On the other hand, the whole system is in $|\Phi^{-}_{p}\rangle|\Phi^{+}_{t}\rangle$ or $|\Psi^{-}_{p}\rangle|\Phi^{+}_{t}\rangle$ with an equal probability of $F_{3}p_{t}$, and in $|\Phi^{-}_{p}\rangle|\Psi^{+}_{t}\rangle$ or $|\Psi^{-}_{p}\rangle|\Psi^{+}_{t}\rangle$ with an equal probability of $F_{3}(1-p_{t})$.

Suppose that the photon pair is in  $|\Phi^{+}_{p}\rangle_{a_{5}b_{5}}|\Phi^{+}_{t}\rangle$.
 As shown in Fig. 1, Alice (Bob) passes the photon in $a_{5}$ $(b_{5})$ through a Pockels cell ($PC_{S}$), which can flip the polarization feature of the incoming photon
under the temporal mode $S$. After the $PC_{s}$, Alice (Bob) passes the photon through a PBS, which makes the photon in $|H\rangle$ be in $a_{7}$ $(b_{7})$ and enter a $PC_{S}$ and the photon in $|V\rangle$ be in $a_{8}$ $(b_{8})$ and enter a $PC_{L}$. As a result, the states $|\Phi^{\pm}_{p}\rangle|\Phi^{+}_{t}\rangle$ will finally evolve to
\begin{eqnarray}
&&|\Phi^{\pm}_{p}\rangle_{a_{3}b_{3}}|\Phi^{+}_{t}\rangle\nonumber\\
&\rightarrow&\frac{1}{2}(|H^{L}H^{L}\rangle_{a_{9}b_{9}}+|V^{S}V^{S}\rangle_{a_{10}b_{10}}\nonumber\\
&\pm&|H^{L}H^{L}\rangle_{a_{10}b_{10}}\pm|V^{S}V^{S}\rangle_{a_{9}b_{9}}).\label{pc1}
\end{eqnarray}
Then, with the help of two PBSs, Alice and Bob can make the photon in $|H\rangle$ pass through the short (S) arm and photon in $|V\rangle$ pass through the long (L) arm. By precisely controlling the length of long and short arms, they can adjust the time-bin feature of the photons in $|H\rangle$ and $|V\rangle$ to be the same. In this way, we can neglect the time-bin features  of the photons, and the state  in Eq. (\ref{pc1}) evolves to
\begin{eqnarray}
&&|\Phi^{\pm}_{p}\rangle_{a_{3}b_{3}}|\Phi^{+}_{t}\rangle\nonumber\\
&\rightarrow&\frac{1}{\sqrt{2}}(|HH\rangle\pm|VV\rangle)\otimes\frac{1}{\sqrt{2}}(|d_{2}d_{6}\rangle\pm|d_{1}d_{5}\rangle),
\end{eqnarray}
 and will be detected by the single photon detector $D_{2}D_{6}$ or $D_{1}D_{5}$. In this case, the second purification step is successful. The polarization feature of the photon pair remains to be $|\Phi^{\pm}_{p}\rangle$.

If the initial state is $|\Psi^{\pm}_{p}\rangle|\Psi^{+}_{t}\rangle$, after the above operations, it will evolve to
\begin{eqnarray}
&&|\Psi^{\pm}_{p}\rangle|\Psi^{+}_{t}\rangle\nonumber\\
&\rightarrow&\frac{1}{\sqrt{2}}(|HV\rangle\pm|VH\rangle)\otimes\frac{1}{\sqrt{2}}(|d_{2}d_{6}\rangle\pm|d_{1}d_{5}\rangle),\label{pc2}
\end{eqnarray}
which will also lead to the successful detection results. The state in Eq. (\ref{pc2}) will finally collapse to $|\Psi^{\pm}_{p}\rangle$.

If the initial state is  $|\Phi^{\pm}_{p}\rangle|\Psi^{+}_{t}\rangle$ or $|\Psi^{\pm}_{p}\rangle|\Phi^{+}_{t}\rangle$, after the above operations, Alice and Bob would never obtain the successful measurement results. In detail, after above operations, we can obtain
\begin{eqnarray}
&&|\Phi^{\pm}_{p}\rangle|\Psi^{+}_{t}\rangle\nonumber\\
&\rightarrow&\frac{1}{\sqrt{2}}(|HV\rangle\pm|VH\rangle)\otimes\frac{1}{\sqrt{2}}(|d_{2}d_{5}\rangle\pm|d_{1}d_{6}\rangle),\nonumber\\
&&|\Psi^{\pm}_{p}\rangle|\Phi^{+}_{t}\rangle\nonumber\\
&\rightarrow&\frac{1}{\sqrt{2}}(|HH\rangle\pm|VV\rangle)\otimes\frac{1}{\sqrt{2}}(|d_{2}d_{5}\rangle\pm|d_{1}d_{6}\rangle),
\end{eqnarray}
which makes the single photon detectors $D_{1}D_{6}$ or $D_{2}D_{5}$ each register a single photon. In this case, the second purification step fails.

On the other hand, if the photon state is in $a_{4}b_{4}$ modes, we can obtain when the photon detectors $D_{3}D_{7}$ or $D_{4}D_{8}$ each register one photon, the second purification step is successful.

Therefore, when the second purification step is successful, we can distill a new mixed state in polarization DOF as
\begin{eqnarray}
\rho_{3p}&=&F'_{1}|\Phi^{+}_{p}\rangle\langle\Phi^{+}_{p}|+F'_{2}|\Phi^{-}_{p}\rangle\langle\Phi^{-}_{p}|\nonumber\\
&+&F'_{3}|\Psi^{+}_{p}\rangle\langle\Psi^{+}_{p}|+F'_{4}|\Psi^{-}_{p}\rangle\langle \Psi^{-}_{p}|,\label{final}
\end{eqnarray}
with a success probability of
\begin{eqnarray}
P_{2}&=&(F_{1}+F_{3})p_{t}+(F_{2}+F_{3})(1-p_{t})\nonumber\\
&=&\frac{1-p_{p}-p_{s}p_{t}+4p_{p}p_{s}p_{t}}{2-2p_{p}-p_{s}+4p_{p}p_{s}}.
\end{eqnarray}
The four coefficients in Eq. (\ref{final}) can be written as
\begin{eqnarray}
F'_{1}&=&\frac{F_{1}p_{t}}{P_{2}}=\frac{3p_{p}p_{s}p_{t}}{1-p_{t}p_{s}+p_{p}(4p_{t}p_{s}-1)}, \nonumber\\
 F'_{2}&=&\frac{F_{3}p_{t}}{P_{2}}=\frac{(1-p_{p})(1-p_{s})p_{t}}{1-p_{t}p_{s}+p_{p}(4p_{t}p_{s}-1)},\nonumber\\
F'_{3}&=&\frac{F_{2}(1-p_{t})}{P_{2}}=\frac{(1-p_{p})(1-p_{t})p_{s}}{1-p_{t}p_{s}+p_{p}(4p_{t}p_{s}-1)}, \nonumber\\
 F'_{4}&=&\frac{F_{3}(1-p_{t})}{P_{2}}=\frac{(1-p_{p})(1-p_{s})(1-p_{t})}{1-p_{t}p_{s}+p_{p}(4p_{t}p_{s}-1)}.
\end{eqnarray}

Similarly as the first step, the second step can reduce the rate of $|\Psi^{\pm}_{p}\rangle$ ($F'_{3}<F_{2}$ and $F'_{4}<F_{3}$). It can be calculated that $F'_{1}>F_{1}$ and $F'_{1}>p_{t}$ when $\frac{3p_{p}p_{s}}+p_{p}-1{p_{s}(4p_{p}-1)}>p_{t}>\frac{1}{2}$. Comparing with original mixed state $\rho_{p}$ in Eq. (\ref{initialp}), after two steps of purification, the rates of $|\Psi^{\pm}\rangle$ and $|\Phi^{-}\rangle$ can be all reduced, so that the fidelity of $|\Phi^{+}_{p}\rangle$ can be efficiently increased.

\section{General entanglement purification}
In this section, we consider a general case that after the long-distance transmission in noisy channel, the entanglement in the spatial-mode and time-bin DOFs also degrade to Werner states. In this way, the mixed states in above two DOFs can be written as
\begin{eqnarray}
\rho_{sn}&=&p_{s}|\Phi^{+}_{s}\rangle\langle\Phi^{+}_{s}|+\frac{1-p_{s}}{3}(|\Psi^{+}_{s}\rangle\langle \Psi^{+}_{s}|\nonumber\\
&+&|\Phi^{-}_{s}\rangle\langle \Phi^{-}_{s}|+|\Psi^{-}_{s}\rangle\langle \Psi^{-}_{s}|),\nonumber\\
\rho_{tn}&=&p_{t}|\Phi^{+}_{t}\rangle\langle\Phi^{+}_{t}|+\frac{1-p_{t}}{3}(|\Psi^{+}_{t}\rangle\langle \Psi^{+}_{t}|\nonumber\\
&+&|\Phi^{-}_{t}\rangle\langle \Phi^{-}_{t}|+|\Psi^{-}_{t}\rangle\langle \Psi^{-}_{t}|).\label{initialfs}
\end{eqnarray}
In Eq. (\ref{initialp}) and Eq. (\ref{initialfs}), we also suppose that $p_{p(t,s)}>\frac{1}{2}$.

In the first step, we only consider $\rho_{p}\otimes\rho_{sn}$, which has 16 possible cases, say $|\Phi^{\pm}_{p}\rangle|\Phi^{\pm}_{s}\rangle$, $|\Phi^{\pm}_{p}\rangle|\Psi^{\pm}_{s}\rangle$, $|\Psi^{\pm}_{p}\rangle|\Phi^{\pm}_{s}\rangle$, and $|\Psi^{\pm}_{p}\rangle|\Psi^{\pm}_{s}\rangle$. By passing the photons in $a_{1}a_{2}$ and $b_{1}b_{2}$ modes through the PBSs, we also select the items which make the spatial modes $a_{3}b_{3}$ or $a_{4}b_{4}$ each have one photon. All the 8 initial states $|\Phi^{\pm}_{p}\rangle|\Phi^{\pm}_{s}\rangle$ and $|\Psi^{\pm}_{p}\rangle|\Psi^{\pm}_{s}\rangle$ can lead to the successful cases. As a result, when the first step is successful, the parties can distill a new mixed state in polarization DOF as
 \begin{eqnarray}
 \rho_{1pn}&=&F_{1n}|\Phi^{+}_{p}\rangle\langle\Phi^{+}_{p}|+F_{2n}|\Phi^{-}_{p}\rangle\langle\Phi^{-}_{p}|\nonumber\\
 &+&F_{3n}(|\Psi^{+}_{p}\rangle\langle\Psi^{+}_{p}|+|\Psi^{-}_{p}\rangle\langle \Psi^{-}_{p}|).\label{step1}
 \end{eqnarray}
with a probability of
\begin{eqnarray}
P_{1n}&=&p_{p}p_{s}+\frac{(1-p_{p})(1-p_{s})}{9}\nonumber\\
&+&\frac{p_{p}(1-p_{s})+p_{s}(1-p_{p})}{3}
+\frac{4(1-p_{p})(1-p_{s})}{9}\nonumber\\
&=&\frac{8p_{p}p_{s}-2p_{p}-2p_{s}+5}{9}.
\end{eqnarray}
In $\rho_{1pn}$, the rates of $|\Phi^{\pm}_{p}\rangle$ and $|\Psi^{\pm}_{p}\rangle$ can be written as
\begin{eqnarray}
F_{1n}&=&\frac{p_{p}p_{s}+\frac{(1-p_{p})(1-p_{s})}{9}}{P_{1n}}\nonumber\\
&=&\frac{10p_{p}p_{s}-p_{p}-p_{s}+1}{8p_{p}p_{s}-2p_{p}-2p_{s}+5},\nonumber\\ F_{2n}&=&\frac{p_{p}(1-p_{s})+(1-p_{p})p_{s}}{3P_{1n}}\nonumber\\
&=&\frac{3p_{p}+3p_{s}-6p_{p}p_{s}}{8p_{p}p_{s}-2p_{p}-2p_{s}+5},\nonumber\\
F_{3n}&=&\frac{2(1-p_{p})(1-p_{s})}{9P_{1n}}\nonumber\\
&=&\frac{2(1-p_{p})(1-p_{s})}{8p_{p}p_{s}-2p_{p}-2p_{s}+5}.
\end{eqnarray}

\begin{figure}
\centering
\includegraphics[width=8cm]{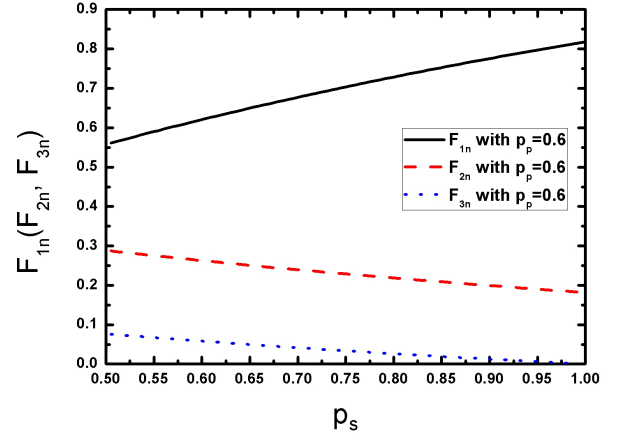}
\caption{The value of $F_{1n}$, $F_{2n}$, and $F_{3n}$ as a function of $p_{s}$. Here, we control $p_{p}=0.6$ and adjust $p_{s}$ from 0.505 to 1. \\ }
\end{figure}

In Fig. 2, we show the values of $F_{1n}$, $F_{2n}$, and $F_{3n}$ as a function of $p_{s}$. Here, we control $p_{p}=0.6$ and adjust $p_{s}$ from 0.505 to 1. It is obvious that both $F_{2n}$ and $F_{3n}$ reduce with the growth of $p_{s}$, which makes $F_{1n}$ increase with the growth of $p_{s}$. We also obtain that $F_{3n}<\frac{1-p_{p}}{3}$ when $p_{s}>\frac{1}{2}$, so that we can reduce the rate of bit-flip error. However, in this general case, as the rate of $|\Phi^{-}_{p}\rangle$ ($F_{2n}$) may be relatively high, and we cannot simply obtain $F_{1n}>p_{p}$ or $F_{1n}>p_{s}$ when $p_{s}>\frac{1}{2}$ and $p_{p}>\frac{1}{2}$. Here, we provide the criterion of $F_{1n}>p_{p}$ as
\begin{eqnarray}
p_{s}>\frac{6p_{p}-2p_{p}^{2}-1}{12p_{p}-8p_{p}^{2}-1},\label{criterion1}
\end{eqnarray}
and
the criterion of $F_{1n}>p_{s}$ as
\begin{eqnarray}
(8p_{p}-2)p_{s}^{2}+6(1-2p_{p})p_{s}+p_{p}-1<0.\label{criterion1n}
\end{eqnarray}

\begin{figure}
\centering
\includegraphics[width=8cm]{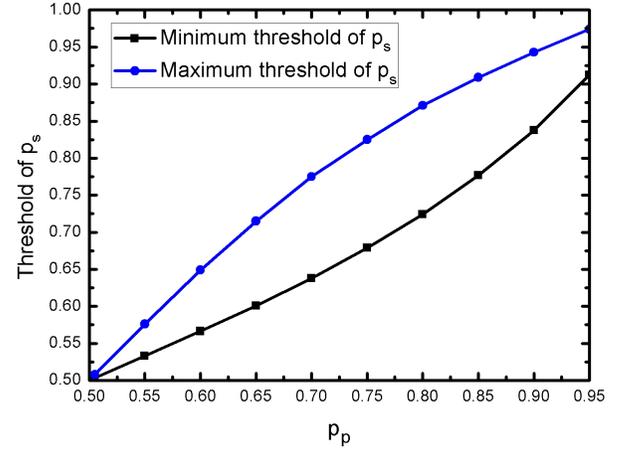}
\caption{The minimum threshold of $p_{s}$ corresponding to $F_{1n}>p_{p}$, and the maximum threshold of $p_{s}$ corresponding to $F_{1n}>p_{s}$ as a function of $p_{p}$. Here, we control $p_{p}\in[0.505,0.95]$. }
\end{figure}
Fig. 3 provides the minimum threshold of $p_{s}$ corresponding to $F_{1n}>p_{p}$, and the maximum threshold of $p_{s}$ corresponding to $F_{1n}>p_{s}$ under different values of $p_{p}$. It can be found that both the minimum and maximum thresholds of $p_{s}$ increase with the growth of $p_{p}$. Combined with Fig. 2 and Fig. 3, we can obtain that a high $p_{s}$ can lead to a small $F_{3n}$ and relatively high $F_{1n}$. However, when the value of $p_{s}$ is too high, we cannot ensure that $F_{1n}>p_{s}$. In this way, for satisfying both the criterions in Eq. (\ref{criterion1}) and Eq. (\ref{criterion2}), the practical value of $p_{s}$ should be between two thresholds. On the other hand, even when $p_{s}$ is relatively high, $F_{2n}$ can be still in a relatively high level, so that we need to perform the second purification step to further reduce $F_{2n}$ and increase the fidelity of $|\Phi^{+}_{p}\rangle$.

In the second purification step, with the help of H operation, we can transform $\rho_{1pn}$ to $\rho_{2pn}$ with the form of
\begin{eqnarray}
\rho_{2pn}&=&F_{1n}|\Phi^{+}_{p}\rangle\langle\Phi^{+}_{p}|+F_{3n}|\Phi^{-}_{p}\rangle\langle\Phi^{-}_{p}|\nonumber\\
&+&F_{2n}|\Psi^{+}_{p}\rangle\langle\Psi^{+}_{p}|+F_{3n}|\Psi^{-}_{p}\rangle\langle \Psi^{-}_{p}|,
\end{eqnarray}
and $F_{2n}$ transforms to the rate of $|\Psi^{+}_{p}\rangle$. In this way, the whole photon system is $\rho_{2pn}\otimes\rho_{tn}$, which also has 16 possible cases. Alice and Bob pass the photons in $a_{3}b_{3}$ and $a_{4}b_{4}$ modes through the purification units. When the detection result is $D_{2}D_{6}$, $D_{1}D_{5}$, $D_{3}D_{7}$ or $D_{4}D_{8}$ each registering one photon, the second purification step will be successful. According to the description in Sec. II, all the states $|\Phi^{\pm}_{p}\rangle|\Phi^{\pm}_{t}\rangle$ and $|\Psi^{\pm}_{p}\rangle|\Psi^{\pm}_{t}\rangle$ can lead to above successful detection. In this way, the success probability of the second step is
\begin{eqnarray}
P_{2n}&=&F_{1n}p_{t}+F_{1n}\frac{1-p_{t}}{3}+F_{3n}p_{t}\nonumber\\
&+&3F_{3n}\frac{1-p_{t}}{3}
+2F_{2n}\frac{1-p_{t}}{3}\nonumber\\
&=&\frac{7-p_{s}+p_{p}(4p_{s}-1)+2p_{t}(4p_{p}-1)(4p_{s}-1)}{3(8p_{p}p_{s}-2p_{p}-2p_{s}+5)}.\nonumber\\
\end{eqnarray}

When the successful detection is obtained, we can distill a new mixed state as
\begin{eqnarray}
 \rho_{3pn}&=&F'_{1n}|\Phi^{+}_{p}\rangle\langle\Phi^{+}_{p}|+F'_{2n}|\Phi^{-}_{p}\rangle\langle\Phi^{-}_{p}|\nonumber\\
 &+&F'_{3n}(|\Psi^{+}_{p}\rangle\langle\Psi^{+}_{p}|+|\Psi^{-}_{p}\rangle\langle \Psi^{-}_{p}|),
 \end{eqnarray}
where
\begin{eqnarray}
F'_{1n}&=&\frac{F_{1n}p_{t}+F_{3n}\frac{1-p_{t}}{3}}{P_{2n}}\nonumber\\
&=&\frac{2(1-p_{p})(1-p_{s})+p_{t}(1-p_{s}-p_{p}+28p_{p}p_{s})}{7-p_{s}-p_{p}+4p_{p}p_{s}+2p_{t}(4p_{p}-1)(4p_{s}-1)},\nonumber\\
F'_{2n}&=&\frac{F_{1n}\frac{1-p_{t}}{3}+F_{3n}p_{t}}{P_{2n}}\nonumber\\
&=&\frac{p_{s}+p_{p}-10p_{p}p_{s}-1+p_{t}(5p_{s}+5p_{p}+4p_{p}p_{s}-5)}{7-p_{s}-p_{p}+4p_{p}p_{s}+2p_{t}(4p_{p}-1)(4p_{s}-1)},\nonumber\\
F'_{3n}&=&\frac{(F_{2n}+F_{3n})\frac{1-p_{t}}{3}}{P_{2n}}\nonumber\\
&=&\frac{(1-p_{t})(p_{s}+p_{p}+2-4p_{p}p_{s})}{7-p_{s}-p_{p}+4p_{p}p_{s}+2p_{t}(4p_{p}-1)(4p_{s}-1)}.\label{step2}
\end{eqnarray}

\begin{figure}
\centering
\includegraphics[width=8cm]{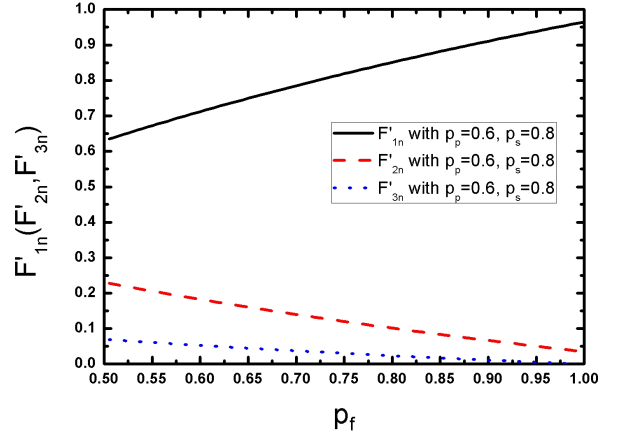}
\caption{The value of $F'_{1n}$, $F'_{2n}$, and $F'_{3n}$ as a function of $p_{t}$. Here, we control $p_{p}=0.6$ and $p_{s}=0.8$, and adjust $p_{t}$ from 0.505 to 1. }
\end{figure}

In Fig. 4, we control $p_{p}=0.6$ and $p_{s}=0.8$ (In this case, we can obtain $F_{1n}\approx0.7285$), and show the values of $F'_{1n}$, $F'_{2n}$, and $F'_{3n}$ as a function of $p_{t}$. It can be found that both $F'_{2n}$ and $F'_{3n}$ reduce with the growth of $p_{t}$, which makes $F'_{1n}$ increase. The higher value of $p_{t}$ leads to higher $F'_{1n}$ and lower $F'_{2n}$ and $F'_{3n}$. It is important to compare $F'_{1n}$ with $F_{1n}$ and $p_{t}$. We can also calculate the criterion for $F'_{1n}>F_{1n}$ as
\begin{eqnarray}
\frac{3(p_{t}-\frac{1}{2})}{1-p_{t}}>\frac{(F_{1n}-\frac{1}{2})(F_{1n}+F_{3n})}{F_{1n}[1-(F_{1n}+F_{3n})]}.\label{criterion2}
\end{eqnarray}
 and the criterion for $F'_{1n}>p_{t}$ as
\begin{eqnarray}
p_{t}^{2}(4p_{p}-1)(4p_{s}-1)+3p_{t}(1-4p_{p}p_{s})<(1-p_{p})(1-p_{s}).\nonumber\\\label{criterion2n}
 \end{eqnarray}

 \begin{figure}
\centering
\includegraphics[width=8cm]{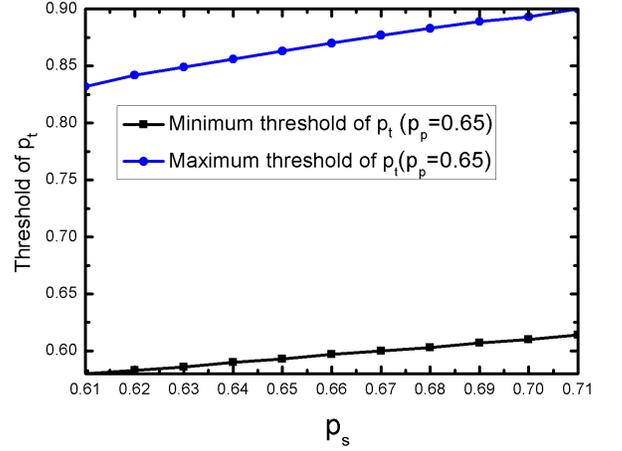}
\caption{The minimum threshold of $p_{t}$ corresponding to $F'_{1n}>F_{1n}$, and the maximum threshold of $p_{t}$ corresponding to $F'_{1n}>p_{t}$ as a function of $p_{s}$. Here, we control $p_{p}=0.65$, and change $p_{s}$ from 0.61 to 0.71. }
\end{figure}

 Similar as the fist step, Eq. (\ref{criterion2}) and Eq. (\ref{criterion2n}) provide the minimum threshold and maximum threshold of $p_{t}$. The practical value of $p_{t}$ should be between the two thresholds. For example, we suppose that $p_{p}=0.65$, where the suitable value of $p_{s}$ should be in the scale $(0.601, 0.715)$. Under this case, we control the $p_{s}$ in the scale of $[0.61, 0.71]$ and provide the minimum value and maximal value of $p_{t}$ altered with the value of $p_{s}$ in Fig. 5.

\section{Discussion}
In the paper, we demonstrate an efficient and simple two-step EPP assisted with hyperentanglement.  This EPP requires only one copy of photon pair hyperentangled in polarization, spatial-mode, and time-bin DOFs. We consider a general degradation model that the entanglement in all DOFs suffer from channel noise. By consuming the imperfect entanglement in spatial-mode and time-bin DOFs, we can reduce both the bit-flip error and phase-flip error in polarization DOF and increase the fidelity of the target polarization state. Our two-step EPP has some attractive advantages. First, comparing with previous recurrence EPPs which require two or more copies of low-quality entangled pairs, our EPP uses the spatial and time-bin information
to complete the measurement pointer, which avoids consuming an entangled photon copy. In this way, our EPP protocol effectively reduces consumption
of entanglement pairs, especially in purification consisting of many rounds. Second, using only one pair of hyperentangled state also reduces the experimental difficulty,
for it is hard to generate two pairs of hyperentangled states simultaneously. Third, all the devices in our EPP are available under current experimental condition, so that our EPP is feasible for experiment.

\begin{figure}
\centering
\includegraphics[width=8cm]{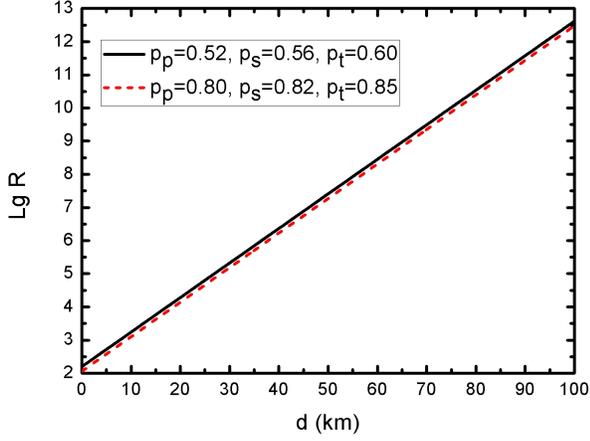}
\caption{The value of $LgR$ as a function of the photon distribution length. We set $\eta_{d}=0.9$ and $\eta_{c}=0.95$ \cite{L}, and consider the low initial fidelity case ($p_{p}=0.52$, $p_{s}=0.56$, $p_{t}=0.60$) and high initial fidelity case ($p_{p}=0.8$, $p_{s}=0.82$, $p_{t}=0.85$), respectively. }
\end{figure}

It is important to compare the purification efficiency of our two-step EPP with previous recurrence EPPs \cite{Pan1} in linear optics in a practical environment. In previous EPPs, for reducing both the bit-flip and phase-flip error in polarization DOF, four pairs of identical low-quality mixed states should be distributed to Alice and Bob (Suppose that the low quality mixed states are the Werner states with the form of Eq. (\ref{initialp})). The transmission efficiency of each photon is $\eta_{t}=e^{-\frac{d}{d_{0}}}$, where $d_{0}$ is the attenuation length of the channel (25 km for commercial fibre \cite{L}) and $d$ is the practical photon transmission distance. They first perform the purification operation on each two identical pairs to reduce the bit-flip error. Only when both the purification operations are successful, they perform the H operation on the distilled two photon pairs and further correct the phase-flip error. We suppose that $\eta_{d}$ and $\eta_{c}$ are the detection efficiency of the practical photon detector and the coupling efficiency of a photon to the photon detector, respectively. The total purification efficiency of previous EPP \cite{Pan1} can be calculated as
\begin{eqnarray}
E_{o}=\frac{1}{4}P_{1t}^{2}P_{2t}\eta_{t}^{8}\eta_{d}^{8}\eta_{c}^{8}.
\end{eqnarray}
Here, $P_{1t}$ and $P_{2t}$ represent the success probability of the first and second purification rounds, respectively, which can be calculated as
\begin{eqnarray}
P_{1t}&=&\frac{1}{2}[p_{p}^{2}+\frac{2p_{p}(1-p_{p})}{3}+\frac{5(1-p_{p})^{2}}{9}],\nonumber\\
P_{2t}&=&\frac{1}{8P_{1t}^{2}}\{[p_{p}^{2}+\frac{(1-p_{p})^{2}}{9}]^{2}+\frac{8}{81}(1-p_{p})^{4}\nonumber\\
&+&\frac{4}{9}[p_{p}^{2}+\frac{(1-p_{p})^{2}}{9}](1-p_{p})^{2}+\frac{4}{9}p_{p}^{2}(1-p_{p})^{2}\nonumber\\
&+& \frac{8}{27}p_{p} (1-p_{p})^{3} \}.
\end{eqnarray}

On the other hand, according to above description, the total purification efficiency of our two-step EPP is
 \begin{eqnarray}
E_{n}=P_{1n}P_{2n}\eta_{t}^{2}\eta_{d}^{2}\eta_{c}^{2}.
\end{eqnarray}
In this way, the ratio of $E_{n}$ and $E_{o}$ can be defined as
\begin{eqnarray}
R=\frac{E_{n}}{E_{o}}=\frac{4P_{1n}P_{2n}}{P_{1t}^{2}P_{2t}\eta_{t}^{6}\eta_{d}^{6}\eta_{c}^{6}}
\end{eqnarray}

Fig. 6 shows the value of $Lg R$ as a function of the photon transmission length $d$. Here, we set $d_{0}=25$ $km$, $\eta_{d}=0.9$ and $\eta_{c}=0.95$ \cite{L}, and change the distance $d$ from 0 to 100 $km$. Here, we select the low initial fidelity case ($p_{p}=0.52$, $p_{s}=0.56$, $p_{t}=0.60$) and high initial fidelity case ($p_{p}=0.8$, $p_{s}=0.82$, $p_{t}=0.85$), respectively. It can be found that the influences from the initial fidelities in three DOFs on $Lg R$ are slight, and $Lg R$ increases linearly with the growth of $d$. In this way, our EPP is extremely useful in the long-distance entanglement distribution.

Next, we discuss the residual entanglement when the purification steps fail. Here, we consider the case in Sec. II for simplicity. We first consider the case that the first purification step fails, but the second step is successful.  When the first step fails, say, the spatial modes $a_{3}b_{4}$ or $a_{4}b_{3}$ each have a photon, the parties can distill a new mixed state with the form of
\begin{eqnarray}
\rho_{fail1}&=&F_{fail1}|\Phi^{+}_{p}\rangle\langle\Phi^{+}_{p}|+A_{fail1}|\Phi^{-}_{p}\rangle\langle\Phi^{-}_{p}|\nonumber\\
 &+&B_{fail1}(|\Psi^{+}_{p}\rangle\langle\Psi^{+}_{p}|+|\Psi^{-}_{p}\rangle\langle \Psi^{-}_{p}|),\label{1f}
\end{eqnarray}
with the probability of
\begin{eqnarray}
P_{fail1}&=&p_{p}(1-p_{s})+\frac{(1-p_{p})(1-p_{s})}{3}+\frac{2(1-p_{p})p_{s}}{3}\nonumber\\
&=&\frac{1+p_{s}+2p_{p}-4p_{p}p_{s}}{3}.\label{pfail1}
\end{eqnarray}
The three coefficients can be calculated as
\begin{eqnarray}
F_{fail1}&=&\frac{p_{p}(1-p_{s})}{P_{fail1}}=\frac{3p_{p}(1-p_{s})}{1+p_{s}+2p_{p}-4p_{p}p_{s}},\nonumber\\
A_{fail1}&=&\frac{\frac{(1-p_{p})(1-p_{s})}{3}}{P_{fail1}}=\frac{(1-p_{p})(1-p_{s})}{1+p_{s}+2p_{p}-4p_{p}p_{s}},\nonumber\\
B_{fail1}&=&\frac{\frac{(1-p_{p})p_{s}}{3}}{P_{fail1}}=\frac{(1-p_{p})p_{s}}{1+p_{s}+2p_{p}-4p_{p}p_{s}}.
\end{eqnarray}
In order to make the distilled mixed state have residual entanglement, we require $F_{fail1}>\frac{1}{2}$. This requirement can be satisfied when $p_{s}<\frac{4p_{p}-1}{1+2p_{p}}$. Under this case, when the second purification step is successful, say, the photon detectors $D_{2}D_{7}$, $D_{1}D_{8}$, $D_{3}D_{6}$, or $D_{4}D_{5}$ each registering a single photon, the parties can obtain a new mixed state as
\begin{eqnarray}
\rho'_{fail1}&=&F'_{fail1}|\Phi^{+}_{p}\rangle\langle\Phi^{+}_{p}|+A'_{fail1}|\Phi^{-}_{p}\rangle\langle\Phi^{-}_{p}|\nonumber\\
 &+&B'_{fail1}|\Psi^{+}_{p}\rangle\langle\Psi^{+}_{p}|+C'_{fail1}|\Psi^{-}_{p}\rangle\langle \Psi^{-}_{p}|,\label{1f2s}
\end{eqnarray}
with the probability of
\begin{eqnarray}
P'_{fail1}&=&(F_{fail1}+B_{fail1})p_{t}+(A_{fail1}+B_{fail1})(1-p_{t})\nonumber\\
&=&\frac{1-p_{p}+p_{t}(4p_{p}-1)(1-p_{s})}{1+p_{s}-2p_{p}(2p_{s}-1)}.
\end{eqnarray}
The coefficients in Eq. (\ref{1f2s}) can be written as
\begin{eqnarray}
F'_{fail1}&=&\frac{F_{fail1}p_{t}}{P'_{fail1}}\nonumber\\
&=&\frac{3p_{t}p_{p}(1-p_{s})}{1-p_{p}+p_{t}(4p_{p}-1)(1-p_{s})},\nonumber\\
A'_{fail1}&=&\frac{B_{fail1}p_{t}}{P'_{fail1}}\nonumber\\
&=&\frac{p_{t}p_{s}(1-p_{p})}{1-p_{p}+p_{t}(4p_{p}-1)(1-p_{s})},\nonumber\\
B'_{fail1}&=&\frac{A_{fail1}(1-p_{t})}{P'_{fail1}}\nonumber\\
&=&\frac{(1-p_{t})(1-p_{s})(1-p_{p})}{1-p_{p}+p_{t}(4p_{p}-1)(1-p_{s})},\nonumber\\
C'_{fail1}&=&\frac{B_{fail1}(1-p_{t})}{P'_{fail1}}\nonumber\\
&=&\frac{p_{s}(1-p_{t})(1-p_{p})}{1-p_{p}+p_{t}(4p_{p}-1)(1-p_{s})}.
\end{eqnarray}
\begin{figure}
\centering
\includegraphics[width=8cm]{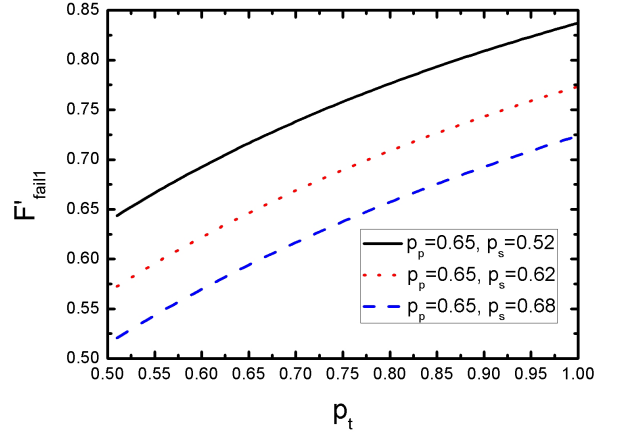}
\caption{The value of $F'_{fail}$ as a function of $p_{t}$. Here, we control $p_{p}=0.65$, so that maximal value of $p_{s}$ for ensuring $F_{fail1}>\frac{1}{2}$ is 0.70. In this way, we control $p_{s}=$ 0.52, 0.62, 0.68, respectively.}
\end{figure}

Fig. 7 shows the value of $F'_{fail1}$ as a function of $p_{t}$ when $p_{p}=0.65$. For ensuring $F_{fail1}>\frac{1}{2}$, we require $p_{s}<0.7$. In this way, we control $p_{s}=$ 0.52, 0.62, 0.68, respectively. It can be found that $F'_{fail1}$ increases with the growth of $p_{t}$, but reduces with the growth of $p_{s}$. With suitable value of $p_{t}$, we can obtain $F'_{fail1}>p_{p}$ and $F'_{fail1}>p_{t}$.

Second, we consider the case that the first purification is successful, but the second purification fails. After the first step, the parties share a mixed state as $\rho_{1p}$ in Eq. (\ref{ro1}). When the second purification step fails, say, the photon detectors $D_{1}D_{6}$, $D_{2}D_{5}$, $D_{3}D_{8}$, or $D_{4}D_{7}$ each registering a single photon, they can distill a new mixed state with the form of
\begin{eqnarray}
\rho_{fail2}&=&F_{fail2}|\Phi^{+}_{p}\rangle\langle\Phi^{+}_{p}|+A_{fail2}|\Phi^{-}_{p}\rangle\langle\Phi^{-}_{p}|\nonumber\\
 &+&B_{fail2}|\Psi^{+}_{p}\rangle\langle\Psi^{+}_{p}|+C_{fail2}|\Psi^{-}_{p}\rangle\langle \Psi^{-}_{p}|,\label{fail2}
\end{eqnarray}
with the probability of
\begin{eqnarray}
P_{fail2}&=&(F_{1}+F_{3})(1-p_{t})+(F_{2}+F_{3})p_{t}\nonumber\\
&=&\frac{1-(1-p_{t})p_{s}-p_{p}+4p_{p}p_{s}(1-p_{t})}{2-p_{s}-2p_{p}(1-2p_{s})}.
\end{eqnarray}
 The coefficients in Eq. (\ref{fail2}) can be written as
\begin{eqnarray}
  F_{fail2}&=&\frac{F_{1}(1-p_{t})}{P_{fail2}}\nonumber\\
  &=&\frac{3(1-p_{t})p_{p}p_{s}}{1-p_{s}+p_{t}p_{s}-p_{p}+4p_{p}p_{s}(1-p_{t})},\nonumber\\
  A_{fail2}&=&\frac{F_{3}(1-p_{t})}{P_{fail2}}\nonumber\\
  &=&\frac{(1-p_{p})(1-p_{s})(1-p_{t})}{1-p_{s}+p_{t}p_{s}-p_{p}+4p_{p}p_{s}(1-p_{t})},\nonumber\\
  B_{fail2}&=&\frac{F_{2}p_{t}}{P_{fail2}},\nonumber\\
  &=&\frac{(1-p_{p})p_{s}p_{t}}{1-p_{s}+p_{t}p_{s}-p_{p}+4p_{p}p_{s}(1-p_{t})},\nonumber\\
  C_{fail2}&=&\frac{F_{3}p_{t}}{P_{fail2}},\nonumber\\
  &=&\frac{(1-p_{p})(1-p_{s})p_{t}}{1-p_{s}+p_{t}p_{s}-p_{p}+4p_{p}p_{s}(1-p_{t})}.
  \end{eqnarray}

\begin{figure}
\centering
\includegraphics[width=8cm]{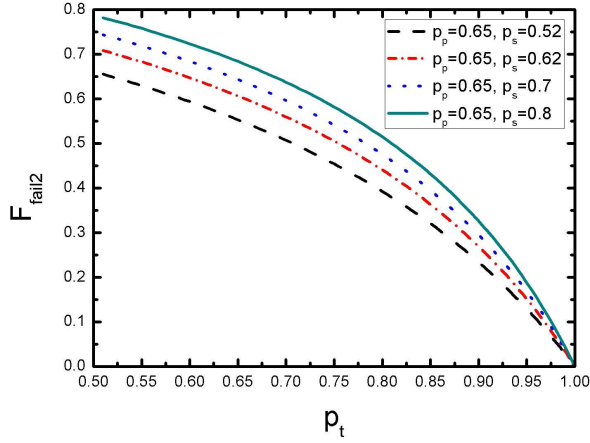}
\caption{The value of $F_{fail2}$ as a function of $p_{t}$. Here, we control $p_{p}=0.65$ and $p_{s}=$ 0.52, 0.62, 0.7, 0.8, respectively, and adjust $p_{t}$ from 0.51 to 1.}
\end{figure}

In Fig. 8, we show the value of $F_{fail2}$ as a function of $p_{t}$ by controlling $p_{p}=0.65$ and $p_{s}=$ 0.52, 0.62, 0.7, 0.8. It can be found that $F_{fail2}$ increases with the growth of $p_{s}$ but reduces with the growth of $p_{t}$. In this way, for obtaining $F_{fail2}>p_{p}$, we require $p_{t}$ to be relatively low. With the growth of $p_{s}$, the maximal values of $p_{t}$ which make $F_{fail2}>p_{p}$ increase. In detail, when $p_{p}=0.65$, and $p_{s}=$ 0.52, 0.62, 0.7, 0.8, the maximal values of $p_{t}$ are 0.519, 0.596, 0.642, and 0.687, respectively.

\begin{figure}
\centering
\includegraphics[width=8cm]{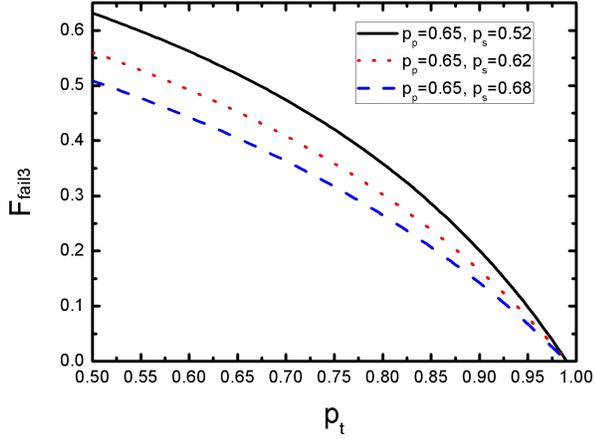}
\caption{The values of $F_{fail3}$ as a function of $p_{t}$. Here, we control $p_{p}=0.65$ and $p_{s}=$ 0.52, 0.62, 0.68, respectively, and adjust $p_{t}$ from 0.51 to 1.}
\end{figure}

Finally, we will discuss the case that both two purification steps fail. This case corresponds to the photon detector $D_{2}D_{8}$, $D_{1}D_{7}$, $D_{3}D_{5}$, or $D_{4}D_{6}$ each registering one photon. After the first purification step, the parties share a new mixed state with the form of $\rho_{fail1}$ in Eq. (\ref{1f}).  Then, when the second purification step fails, they can finally obtain a new mixed state as
\begin{eqnarray}
\rho_{fail3}&=&F_{fail3}|\Phi^{+}_{p}\rangle\langle\Phi^{+}_{p}|+A_{fail3}|\Phi^{-}_{p}\rangle\langle\Phi^{-}_{p}|\nonumber\\
 &+&B_{fail3}|\Psi^{+}_{p}\rangle\langle\Psi^{+}_{p}|+C_{fail3}|\Psi^{-}_{p}\rangle\langle \Psi^{-}_{p}|,\label{fail3}
\end{eqnarray}
with the probability of
\begin{eqnarray}
P_{fail3}&=&(F_{fail1}+B_{fail1})(1-p_{t})+(A_{fail1}+B_{fail1})p_{t},\nonumber\\
&=&\frac{p_{p}(3-4p_{s})-p_{t}(4p_{p}-1)(1-p_{s})+p_{s}}{1-2p_{p}(2p_{s}-1)+p_{s}}.
\end{eqnarray}
The four coefficients can be calculated as
\begin{eqnarray}
  F_{fail3}&=&\frac{F_{fail1}(1-p_{t})}{P_{fail3}}\nonumber\\
  &=&\frac{3p_{p}(1-p_{t})(1-p_{s})}{p_{p}(3-4p_{s})-p_{t}(4p_{p}-1)(1-p_{s})+p_{s}},\nonumber\\
  A_{fail3}&=&\frac{B_{fail1}(1-p_{t})}{P_{fail3}}\nonumber\\
 &=&\frac{(1-p_{p})(1-p_{t})p_{s}}{p_{p}(3-4p_{s})-p_{t}(4p_{p}-1)(1-p_{s})+p_{s}},\nonumber\\
  B_{fail3}&=&\frac{A_{fail1}p_{t}}{P_{fail3}}\nonumber\\
  &=&\frac{(1-p_{p})(1-p_{s})p_{t}}{p_{p}(3-4p_{s})-p_{t}(4p_{p}-1)(1-p_{s})+p_{s}},\nonumber\\
  C_{fail3}&=&\frac{B_{fail1}p_{t}}{P_{fail3}}\nonumber\\
  &=&\frac{(1-p_{p})p_{s}p_{t}}{p_{p}(3-4p_{s})-p_{t}(4p_{p}-1)(1-p_{s})+p_{s}}.
  \end{eqnarray}

In Fig. 9, we show the value of $F_{fail3}$ as a function of $p_{t}$ with $p_{p}=0.65$. For ensuring the existence of residual entanglement after the first purification step, we require $p_{s}<0.7$. In this way, we control $p_{s}=$ 0.52, 0.62, and 0.68, respectively. It can be found that $F_{fail3}$ reduces with the growth of $p_{s}$ and $p_{t}$.  $F_{fail3}$ is lower than $p_{p}$ under arbitrary value of $p_{s}$ and $p_{t}$. However, with relatively low $p_{s}$ and $p_{t}$, there may still exist residual entanglement ($F_{fail3}>\frac{1}{2}$) in  $\rho_{fail3}$. For example, for ensuring the existence of residual entanglement in  $\rho_{fail3}$, we can calculate the maximal values of $p_{t}$ to be 0.682, 0.599, and 0.523, corresponding to $p_{s}=$ 0.52, 0.62, and 0.68, respectively.

From above discussion, there may residual entanglement exist when the purification steps fail. When only one purification step fails, the parties may still distill high-quality entanglement with suitable $p_{s}$ and $p_{t}$ by using of the residual entanglement. Even if the parties can not directly obtain high-quality polarized mixed state, they can still reuse the residual entanglement to distill high-quality entanglement in the next purification round. The existence of residual entanglement provides us a possibility to increase the fidelity of polarization entanglement, thus increase the yield of our EPP.

Finally, it is interesting to compare out two-step EPP with the deterministic EPPs (DEPPs), which also adopt the hyperentanglement to realize the purification \cite{shenghyper1,shenghyper2}. The DEPPs require one pair of hyperentangled state, i.e., polarization-spatial-mode and polarization-spatial-mode-frequency hyperentanglement, respectively. Actually, the DEPPs can completely transform  the entanglement in the other DOF to the target DOF, and they do not require the initial target DOF to be entangled. The upper bound of the fidelity in the target DOF is the initial fidelity of the consumed entanglement.  On the other hand, our current two-step EPP belongs to recurrence EPP. In our two-step EPP, we suppose that the initial fidelities in three DOFs are larger than $\frac{1}{2}$. After two purification steps, the fidelity of the target polarization state can be higher than $p_{p}$, $p_{s}$, and even $p_{t}$.
Actually, in the first purification step, if $p_{p}<\frac{1}{2}$, we can obtain the fidelity $p_{p}<F_{1n}<p_{s}$. Under this case, when the $p_{p}$ and $p_{s}$ satisfy $p_{p}p_{s}>\frac{1}{4}$, we can obtain $F_{1n}>\frac{1}{2}$, say, the distilled new mixed state in polarization DOF has entanglement. In the second purification step, the fidelity of the target polarization DOF can be increased to be higher than $F_{1n}$ and even $p_{t}$. If $p_{p}$ is so low that $p_{p}p_{s}>\frac{1}{4}$ can not be satisfied, our two-step EPP can not work. Based on above comparison, if the initial fidelity in the polarization DOF is relatively high, the current two-step EPP may be more advantageous, and while if that of the polarization DOF is low, the DEPP may be more advantageous.

\section {Conclusion}
In conclusion, we present an efficient two-step recurrence EPP for purifying the entanglement in polarization DOF. In the protocol, we only require one copy of two-photon pair, which is hyperentangled in polarization, spatial-mode, and time-bin DOFs. We suppose that after the photon transmission, the entanglement in all DOFs suffer from the channel noise and degrade to mixed states. As the entanglement in different DOFs have different noisy robustness, the initial mixed states in three DOFs have different fidelities. In the first purification step,  the bit-flip error in polarization DOF can be reduced by consuming the imperfect spatial-mode entanglement, while in the second step, the phase-flip error in polarization DOF can be reduced by consuming the imperfect time-bin entanglement. As a result, the fidelity of the target polarization state can be efficiently increased. Our EPP has some attractive advantages. First, comparing with previous two-step recurrence EPPs, which require two or more same copies of nonlocal entangled pairs, our EPP largely reduces the consumption of entanglement pairs. Second, using only one pair of hyperentangled state also reduces the experimental difficulty,
for it is hard to generate two pairs of hyperentangled states simultaneously. Third, if we consider the practical photon transmission and detector efficiency, our EPP has much higher purification efficiency. Forth, all the devices in our EPP are available under current experimental condition, so that our EPP is feasible for experiment.
 Moreover, in traditional two-step recurrent EPP, the parties can distill a high-quality entanglement only when both two steps are successful. If any one step fails, there are no residual entanglement in the distilled mixed state and the distilled photon states have to be discarded. However, when a purification step of our EPP fails, there may exist residual entanglement in the distilled mixed state. The existence of residual entanglement may make the parties distill higher-fidelity polarization entanglement. Even not, the residual entanglement  may be reused in the next purification round. In this way, the existence of residual entanglement benefits for further increasing the yield of our EPP.  All the above features
make our EPP protocol have potential application in future
quantum information processing field.

\section*{ACKNOWLEDGEMENTS}
This work was supported by  the National Natural Science  Foundation of China (No. 11974189).

\end{document}